# Dark Matter and Charged Exotic Dust

Gerald E. Marsh

Argonne National Laboratory (Ret)

5433 East View Park

Chicago, IL 60615

E-mail: gemarsh@uchicago.edu

**Abstract.** The density profiles of dark matter halos are often modeled by an approximate solution to the isothermal Lane-Emden equation with suitable boundary conditions at the origin. It is shown here that such a model corresponds to an exact solution of the Einstein-Maxwell equations for exotic charged dust. It is also shown that, because of its necessarily very small charge to mass ratio, the fact that the particles are charged does not necessarily rule out such material as a candidate for dark matter.

**Introduction**

The standard ΛCDM has as its principal matter component collisionless cold dark matter of an unknown nature. The rotation curves of spiral galaxies as well as the inferred mass of galaxy clusters are best explained by the existence of dark matter that dominates their mass content. Relatively recent work on colliding galaxy clusters appear to confirm this supposition.[1,2] Other possibilities, such as modifying Newton's equations or postulating a change in the gravitational interaction between dark and normal matter have been proposed, but have not gained favor.

In the case of the rotation curves of galaxies, the density distribution of dark matter is generally assumed to be spherical and to have an isothermal equation of state; i.e., a polytropic equation of state ($P = K\rho^\gamma$) where $\gamma = 1$. The hydrostatic balance equation may then be integrated to yield

$$\rho = \rho_0 \exp\left(-\frac{\Phi}{K}\right), \qquad (1)$$

where $\Phi$ is the gravitational potential. $\Phi/K$ must then be a solution of the isothermal Lane-Emden equation. Non-singular solutions can be obtained by imposing appropriate boundary conditions, such as requiring that the solution and its first derivative vanish at the origin. The result is an exponential solution for the density of the form

$$\rho(r) = \frac{\rho_0}{\exp\left(\frac{\Phi}{K}\right)}. \qquad (2)$$

The isothermal Lane-Emden equation cannot be solved analytically and consequently $\Phi/K$ is expanded in a power series. The requirement that the first derivative vanish at the origin limits the expansion to even powers starting with $(\Phi/K)^2$. Expanding the exponential in the denominator of Eq. (2), keeping only the first two terms, and using the coefficient given by Chandrasekhar[3] for the leading $(\Phi/K)^2$ term results in the often used expression for the dark matter density,

$$\rho(r) \approx \rho_0 \frac{r_0^2}{r_0^2 + r^2}, \qquad (3)$$

where $r_0 = \sqrt{\dfrac{6K}{4G\rho_0}}$. It will be seen that the right hand side of this approximate expression corresponds to an *exact* solution of the coupled Einstein-Maxwell equations for exotic charged dust. Note that if $r_0$ is to be identified with the King radius, the numerical factor of 6 should be replaced by 9.

**Charged Dust**

The form of the metric for charged dust was introduced by Majumdar[4] and Papapetrou[5]. It is spherically symmetric and static, and can be motivated by considering the Reissner-Nordström metric

$$ds^2 = \left(1 - \frac{2m}{r} + \frac{Q^2}{r^2}\right)dt^2 - \left(1 - \frac{2m}{r} + \frac{Q^2}{r^2}\right)^{-1} dr^2 - dr^2(d\theta^2 + \sin^2\theta\, d\phi^2). \tag{4}$$

Assume the extreme form of this metric where $|Q| = m$, and introduce the isotropic coordinates $\bar{r} = r - m$. Doing so results in the metric

$$ds^2 = f^2 dt^2 - f^{-2}\left[d\bar{r}^2 + \bar{r}^2(d\theta^2 + \sin^2\theta\, d\phi^2)\right], \tag{5}$$

where $f = \left(1 + \dfrac{m}{\bar{r}}\right)^{-1}$. Henceforth the bar above the $r$ will be dropped with the understanding that isotropic coordinates are used in what follows.

Using Newtonian mechanics and classical electrostatics, it is straightforward to show that a system of charged particles of mass $m_i$ and charge $e_i$, where all of the particles have the same sign charge, will be in static equilibrium if $|e_i| = G^{1/2} m_i$. For a continuous distribution of mass $\rho$ and charge $\sigma$, there will be equilibrium everywhere if $|\sigma| = G^{1/2}\rho$. This is what is known as charged dust. It has a general relativistic analog that was discovered by Papapetrou and Majumdar. Although spatial symmetry is not required, spherical symmetry will be assumed here.

The equilibrium of charged dust in general relativity has been treated extensively by W.B. Bonnor and others since the early 1960s. It is his paper on the equilibrium of a charged sphere[6] that forms the embarkation point for the work here.[7] The Einstein and Maxwell field equations applied to the metric of Eq. (5) show that the Newtonian condition for equilibrium given above must also hold in general relativity. In what follows, the charge will be chosen to be positive.

Bonnor obtained the equation that relates the general form of *f* to the density,

$$ff'' - 2f'^2 + \frac{2}{r} ff' - 4\kappa\rho = 0. \tag{6}$$

Unfortunately, this equation is completely intractable unless $\rho = 0$, and as put by Lemos and Zanchin, "It is not a method for solving the differential equation of the Majundar-Papapetrou problem, it is an art of correct guessing."[8] In other words, one is reduced to guessing a form for the function *f* and hoping that the equation yields a physically meaningful density distribution.

The problem addressed by Bonnor was to find the density distribution of charged dust within a finite sphere of radius *a* that would match to the vacuum Reissner-Nordström solution at the boundary. This was successfully achieved using the following expression for *f*

$$f(r) = \left(a^3 + mr^2\right)^{\frac{1}{2}}(a+m)^{-\frac{3}{2}}. \tag{7}$$

In Eq. (7), *m* is the mass of the charged dust contained within $r = a$. The density was found to be

$$\rho = \frac{3m}{4\pi a^3}\left(1+\frac{m}{a}\right)^{-3}\left(1+\frac{mr^2}{a^3}\right)^{-1}. \tag{8}$$

The question addressed here is whether it is possible to find a function $f(r)$ that would result in a radially unlimited density distribution matching that given in Eq. (3) for dark matter. Indeed one can. Substitution of

$$f(r) = \sqrt{\frac{4\kappa}{3}\rho_0}\left(a^2 + r^2\right)^{\frac{1}{2}} \tag{9}$$

into Eq (6) yields

$$\rho(r) = \rho_0 \frac{a^2}{a^2 + r^2}, \tag{10}$$

where *a* is now a free constant. This has the same form as Eq. (3) except that now the equality is exact and $\rho(r)$ is derived from a solution of the Einstein-Maxwell field equations. This is somewhat surprising given that the origins to Eq. (3) and Eqs. (9) and (10) are so different.

There are also other solutions to Eq. (6) for the density distribution that exhibit unusual profiles that could prove useful. They result from the form of $f$ given by

$$f(r) = (a + br^n)^{\frac{1}{2}}\left(1 - \frac{1}{1 + cr}\right), \quad n = 1, 3/2 \tag{11}$$

for certain ranges of the constants $a$, $b$, and $c$. Figure 1 shows a density plot for these solutions.

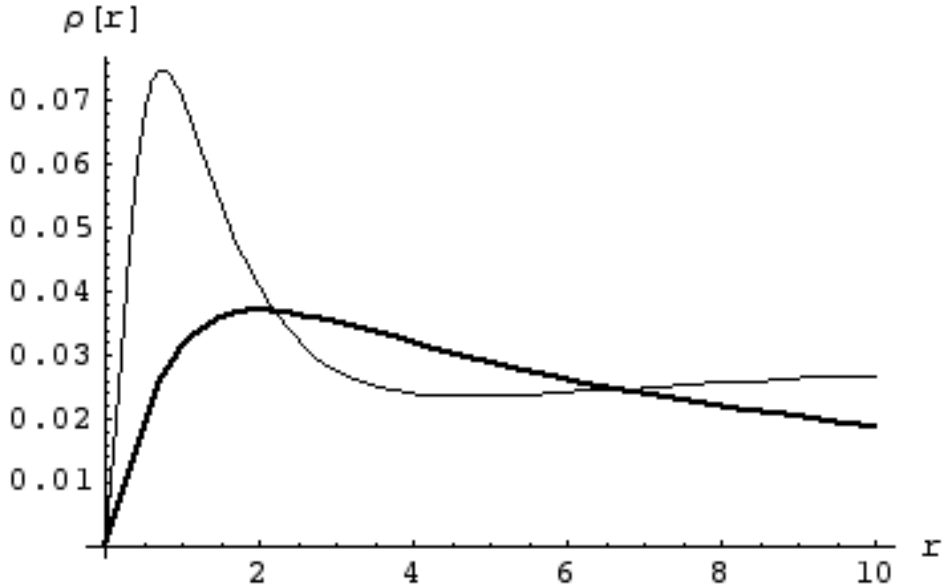

Figure 1. A plot of the density distribution as a function of $r$ given by Eq. (11) for $n = 1$ (dark) and $n = 3/2$ (light) for $a = b = c = 1$.

Note that for $n = 3/2$ the density vanishes at $r = 0$, peaks, and then rises again (peaking to ~ 0.028 at $r \sim 15$) before it decreases slowly to zero at infinity. There is observational support for such dark energy distributions—where, however, the density does not vanish at the origin—in galactic clusters.[9] Other solutions to Eq. (6) may exist that more closely match this observation. For $n = 1$, on the other hand, after the initial peak the density decreases monotonically.

If one tries to generalize the solution of Eq. (9) to

$$f(r) = \sqrt{\frac{4}{3}\rho_0}(a^\alpha + r^\alpha)^{\frac{1}{2}}, \tag{12}$$

the solution for the density is found to be

$$\rho(r) = \rho_0 \frac{r^{\alpha-2}[2a^\alpha(1+\alpha) - r^\alpha(\alpha-2)]}{12(a^\alpha + r^\alpha)}. \tag{13}$$

For $\alpha > 2$, the density has negative values for some range of $r$. Consequently, one must impose the condition that $\alpha \leq 2$. The plot of the density for $\alpha = 2$ and $\alpha < 2$ is shown in Fig. 2. The cusp in the density for $\alpha < 2$ is a result of the density being singular at $r = 0$. Singular density functions are often used to model the mass distributions in elliptical galaxies. This is possible because the total mass as a function of $r$ is finite.[10] This also the case for the density function given by Eq. (13).

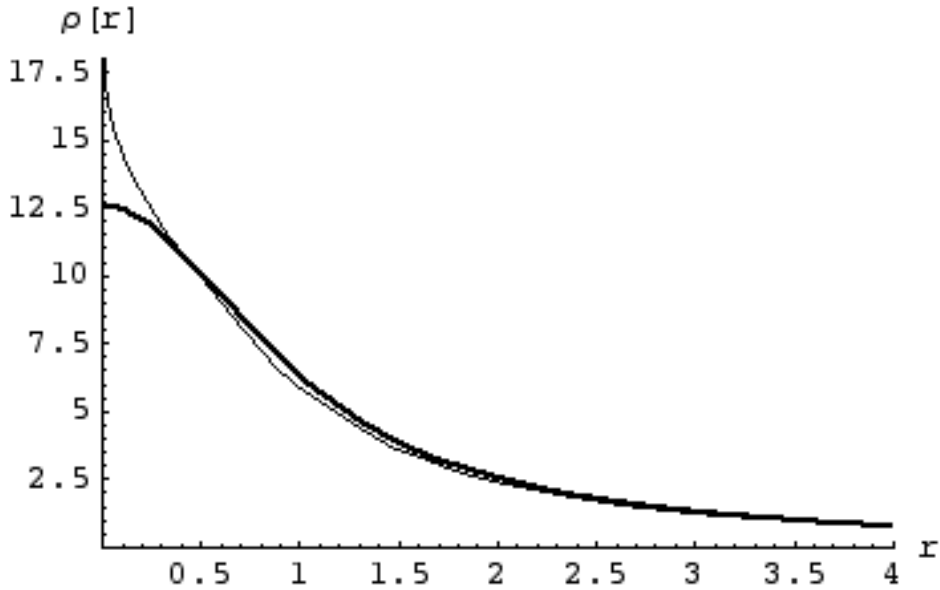

Figure 2. A plot of the density distribution as a function of $r$ given by Eq. (13) for $\alpha = 2$ (dark) and $\alpha = 1.9$ (light) for $\rho_0 = a = 1$. The density for $\alpha = 1.9$ is singular at $r = 0$.

Note that in computing the total mass the gravitational potential energy must also be included. This implies that one does not use the proper volume element in carrying out the integration of the density; i.e.,[11]

$$M(r) = 4\pi \int_0^r \frac{r^{-2}[2a(1+\alpha) - r(\alpha-2)]}{12(a+r)} r^2 dr = \frac{1}{3}\pi r$$

$$\times \left(3\alpha a - \frac{r(\alpha-2)}{\alpha+1} - 3\alpha a\, Hypergeometric2F1\left[\frac{1}{\alpha}, 1, 1+\frac{1}{\alpha}, -a^{-\alpha} r\right]\right). \quad (14)$$

Given the result above for the density distribution, it is not surprising that for $\alpha > 2$, the total mass inside a radius $r$ becomes negative as $r$ increases, so that $\alpha$ must again be limited to values less than or equal to 2.

**Charged Exotic Dust as Dark Matter**

Use of the term "charged dust" here is justified by the long history of the term in the field of general relativity and the extensive work done by Bonnor and others mentioned earlier. If charged dust is to be considered at all as a possible candidate for dark matter, in cannot be composed of particles of conventional matter—for reasons given below, hence the "exotic" qualifier.

The extremal condition $q = G^{1/2} m$ means that if $q$ is chosen to be the minimal charge of one electron or $10^{-19}$ coulomb, then there is a minimal mass of $\sim 3.6 \times 10^{-9}$ kilogram giving a charge to mass ratio of $2.7 \times 10^{-11}$. This minimal mass is unusual in that it is very close in value to the reduced Plank mass of $\sqrt{\frac{\hbar c}{8\pi G}} = 4.3 \times 10^{-9}$ kilogram, and is much greater than the supersymmetric extension of the standard model predicting WIMPs having a mass of $\sim 100$ Gev/c$^2$. In what follows, the charged exotic dust constituents should be thought of as unknown, heavy particles carrying a minimal charge.

It is the fact that Eq. (10), giving an exact solution of the coupled Einstein-Maxwell equations for the density, matches the usual model for galactic dark matter density that gives the incentive to consider charged exotic dust as a possible dark matter candidate. At first, the possibility seems absurd: the fact that the particles making up the dust are charged would seem to rule it out for dark matter because electromagnetic interactions would give the particles too high of a cross section. But, as will be shown, this is not the case.

Treated as a free charged particle, the Thomson scattering cross section for charged exotic dust to electromagnetic radiation, which goes as $\left(\frac{e^2}{4\pi\epsilon_0 mc^2}\right)^2$, would be $\sim 10^{-72}$ m$^2$, so that Thomson scattering would not rule out charged exotic dust as dark matter.

Coulomb scattering is somewhat more complicated to estimate due to the inability to obtain a finite expression for the total cross section. This is, of course, due to the infinite range of the coulomb force that results in the total number of particles scattered through some angle, no matter how small, being infinite. Note that the Rutherford formula, as generally written, assumes the particles scattered to have an incoming density of 1 per cm$^2$ (cgs units will be used in discussing coulomb scattering). One way to get a reasonable expression for the total cross section is to introduce a minimum scattering angle $\theta_{min}$, and this is what will be done here.

The velocity associated with the galaxy cluster collisions of the Bullet Nebula and MACS system is $\sim 4 \times 10^8$ cm/sec. With respect the rest frame of the charged exotic dust particles, this means that the kinetic energy of a typical particle is $\sim 10^{-6}$ ergs (90% of intergalactic gas density is due to H$_2$, which will be used here and assumed to be singly ionized). The amount of energy transferred to a charged exotic dust particle by elastic scattering depends on the scattering angle and is given by

$$E^{xfr} = \frac{2\, m_1^2\, m_2\, v_1^2}{(m_1 + m_2)^2} \sin^2\frac{\theta}{2},$$

(15)

where $m_1$ is the mass of the H$_2$ molecule, $m_2$ is the exotic dust particle mass, $v_1$ is the average H$_2$ molecule velocity, and $\theta$ is the scattering angle. The *fraction* of the original H$_2$ kinetic energy transferred to the dust particle is

$$E(\theta) = \frac{4\, m_1\, m_2}{(m_1 + m_2)^2} \sin^2\frac{\theta}{2}.$$

(16)

---

See references 1 and 2.

The Rutherford scattering formula, giving the number of $H_2$ molecules per unit solid angle scattered to an angle $\theta$, assuming an incoming scattered particle density of one molecule per $cm^2$, is

$$\frac{d\sigma}{d\Omega} = \frac{Z^2 e^4}{16 E^2} \sin^{-4}\frac{\theta}{2} \sim 10^{-26} \sin^{-4}\frac{\theta}{2}. \tag{17}$$

Now the average number density in intergalactic space is $\sim 10^{-3}/cm^3$, so that the areal density over the distance associated with the Bullet and MACS systems ($\sim 10^{24}$ cm) is about $10^{21}/cm^2$. To compute the total energy transferred over this distance, the differential cross section given in Eq. (17) must be multiplied by this value giving

$$\frac{d\sigma}{d\Omega} = 10^{-5} \sin^{-4}\frac{\theta}{2}. \tag{18}$$

The total energy transferred by scattering to an angle $\theta$ is then the areal density times the product of $\Delta E(\theta)$ and $\frac{d\sigma}{d\Omega}$ as given in Eq. (18). Here $m_1 \ll m_2$, so that the total energy transferred, integrated over all angles, is

$$E_{tot}^{xfr} = 8\pi \times 10^{-22} \int_{\theta_{min}} \sin^{-2}\frac{\theta}{2} \sin\theta \, d\theta. \tag{19}$$

This may be written in terms of the impact parameter associated with $\theta_{min}$, $b_{min} = \frac{e^2}{2E}\cot\frac{\theta_{min}}{2}$, as

$$E_{tot}^{xfr} = 8\pi \times 10^{-22} \int_{2\cot^{-1}(10^{13}b_{min})} \sin^{-2}\frac{\theta}{2} \sin\theta \, d\theta. \tag{20}$$

As expected, this integral diverges for $b_{min} = \infty$, but for any reasonable value of $b_{min}$, the total energy transferred is around $10^{-19}$ erg. This result means that Coulomb scattering also does not rule out charged exotic dust as a candidate for dark matter.

**Self-Interaction of charged exotic dust**

In discussing "collisionality" for dark matter, one often uses the measure $\sigma/m$, the cross section per unit mass otherwise known as the opacity. There now exist limits on this ratio from the galactic cluster collisions of the Bullet Nebula and the MACS system, both referred to earlier. The cross section per unit mass for the self interaction of dark matter is set by these collisions at $\sigma/m \lesssim 1$. For charged exotic dust, $\sigma/m = 1$ corresponds to a cross section of $\sigma \sim 10^{-9}$ m². If charged exotic dust is to be considered as a candidate for dark matter, this cross section cannot correspond to a *geometrical* cross section for its interaction with intergalactic gas, which has a number density ~10/m³, because the mean free path would be far too short over the galactic cluster scale.

The *local* dark matter density has been estimated to be[12] $\rho_0 = 0.4$ GeV/cm³, or about one proton mass per cm³. For the charged exotic dust mass, the number density would then be ~$10^{-18}$/cm³. From the mean free path $\lambda = (\sigma \rho_n)^{-1}$ and the Bullet and MACS system length scale, one can see that the cross section is $\sigma \sim 10^{-6}$ cm² giving $\sigma/m$~1. This is at the high end since the local value for the dark matter density has been used rather than that for intergalactic space, which would be expected to be smaller.

Thus, there appears to be no compelling reason to rule out charged exotic dust as a candidate for dark matter. And, as mentioned earlier, unlike the usual model for dark matter as an isothermal sphere, charged exotic dust is essentially static. This would make detection difficult since the galaxy's visible matter and dark matter co-rotate.

**Summary**
Charged exotic dust is an unusual and counter intuitive candidate for dark matter. What makes it worth considering is that some of the solutions to the coupled Einstein-Maxwell equations for its density profiles match the usual density models used for dark matter. The minimal charge for charged exotic dust implies a very large particle mass that is very close to matching the Planck mass. It would be interesting to understand if there is any significance to this or whether it is simply a coincidence. The very small charge to mass ratio for exotic charged dust particles means that Coulomb and Thomson scattering do not rule it out as a candidate for dark matter. The limits on dark matter density, and the large mass of minimal charged exotic dust particles, also tells us that the self-interaction cross section is very small and easily satisfies the constraint of $\sigma/m \lesssim 1$ imposed by the Bullet and MACS system galactic cluster collisions.

# REFERENCES

[1] D. Clowe, et al., A Direct Empirical Proof of the Existence of Dark Matter", Astrophysical Journal, 648:L109–L113, 2006.

[2] M. Brada_, S. W. Allen, and S. W. Allen, "Revealing the properties of dark matter in the merging cluster MACS J0025.4-1222," Astrophysical Journal, 687: 959–967, 2008.

[3] S. Chandrasekhar, *An Introduction to Stellar Structure* (Dover Publications, Inc., University of Chicago 1939).

[4] S.D. Majumdar, "A Class of Exact Solutions of Einstein's Field Equations", *Phys. Rev.* **72**, 390-398 (1947).

[5] A. Papapetrou, *Proc. R. Irish Acad.*, **51**, 191 (1947).

[6] W.B. Bonnor, "The Equilibrium of a Charged Sphere", *Mon. Not. R. Astron. Soc.* **129**, 443-446 (1965).

[7] Calculations of the Einstein and other tensors was performed using the EDCRGTCcode hosted in Mathematica.

[8] J.P.S. Lemos, and V.T. Zanchin, arXiv: gr-qc/0802.0530 v2.

[9] M.J. Jee, et al., Astrophysical Journal, 661: 728-749, 2007.

[10] For example, if $M(r) = r$ , then the density is singular at the origin for  $< 3$.

[11] The fact that the total mass inside a radius *r* must include the gravitational potential energy has been discussed for, for example, by: S.L. Shapiro and S.A. Teukolsky, Black Holes, White Dwarfs and Neutron Stars (WILEY-VCH Verlag GmbH & Co. KGaA, Weinheim 2004). Discussion may also be found in the context of the Tolman-Oppenheimer-Volkov equation.

[12] N. Bernal, "Constraining Dark Matter Properties and the Milky Way Dark Matter Density Profile with Fermi-LAT", Avignon, April 21, 2011 (http://www.th.physik.uni-bonn.de/People/nicolas/cosas/potsdam11.pdf).